\documentclass{PoS}

\usepackage{subfigure}

\leftline{DESY 09-167}
\leftline{Liverpool LTH 847}

\title{Wilson loops in very high order lattice perturbation theory}

\ShortTitle{Wilson loops at very high order LPT}

\author{E.-M.~Ilgenfritz{$^1$}, Y.~Nakamura{$^2$}, \speaker{H.~Perlt}{$^3$}, P.E.L.~Rakow{$^4$}, G.~Schierholz{$^{2,5}$} and A.~Schiller{$^3$}\\
        \llap{$^1$}Institut f\"ur Theoretische Physik, Ruprecht-Karls-Universit\"at Heidelberg, Philosophenweg 16,\\
D-69120 Heidelberg, Germany\\
         \llap{$^2$}Institut f\"ur Theoretische Physik, Universit\"at Regensburg, Universitätsstr. 31,\\
D-93053 Regensburg, Germany\\
         \llap{$^3$}Institut f\"ur Theoretische Physik, Universit\"at Leipzig, PF 100 920, 
D-04009 Leipzig, Germany\\
         \llap{$^4$}Theoretical Physics Division, Department of Mathematical Sciences, 
University of Liverpool,Liverpool L69 3BX, UK\\
         \llap{$^5$}DESY, Theory Group, Notkestrasse 85,
D-22603 Hamburg, Germany \\
       E-mail: \email{perlt@itp.uni-leipzig.de}}

\abstract{We calculate Wilson loops of various sizes
up to loop order $n=20$ for lattice sizes of $L^4 (L=4, 6, 8, 12)$
using the technique of Numerical Stochastic
Perturbation Theory in quenched QCD.
This allows to investigate the  behaviour of
the perturbative series at high orders. We discuss three models
to estimate the perturbative series: a renormalon inspired fit, a heuristic fit
based on an assumed power-law singularity and boosted perturbation
theory. We have found differences in the behavior of the perturbative series for smaller and
larger Wilson loops at moderate $n$. A factorial growth of the coefficients could
not be confirmed up to $n=20$. From Monte Carlo measured plaquette data and our perturbative
result we estimate a value of the gluon condensate 
$\langle \frac{\alpha}{\pi}GG \rangle$.
}

\FullConference{The XXVII International Symposium on Lattice Field Theory - LAT2009\\
		 July 26-31 2009\\
		 Peking University, Beijing, China}

\begin{document}

\section{Introduction}

Since the introduction of the non-perturbative
gluon condensate by Shifman, Vainshtein and Zakharov~\cite{Shifman:1978bx} there have been many attempts to obtain 
reliable numerical results for this quantity. 
Soon it became clear that lattice gauge theory provides a promising tool to
calculate it from Wilson loops. In ~\cite{Banks:1981zf}
the plaquette was used whereas larger Wilson loops have been investigated 
in~\cite{Kripfganz:1981ri}.
From the plaquette $P$ the non-perturbative
gluon condensate $\langle \frac{\alpha}{\pi}G\,G \rangle$ is conventionally derived from
the relation
\begin{equation}
P_{MC} = P_{pert} - a^4 \frac{\pi^2}{36} \left[\frac{-b_0\,g^2}{\beta(g)}\right]\langle \frac{\alpha}{\pi}GG \rangle\,,
\label{GGdef}
\end{equation}
where $b_0$ is the first coefficient of the $\beta$-function and $P_{MC}$ is the plaquette 
 measured in Monte Carlo. In (\ref{GGdef}) it is assumed that the non-perturbative
part scales like the fourth power of the lattice spacing $a$. However, there were speculations
that there could be non-perturbative contributions which scale like $a^2$~\cite{Burgio:1997hc}.
In the last decade the application of Numerical Stochastic Perturbation Theory (NSPT)
\cite{Alfieri:2000ce} pushed the perturbative order of $P_{pert}$ up to order $n=10$
\cite{DiRenzo:2000ua} and even $n=16$~\cite{Rakow:2005yn}. This strongly supports
to use (\ref{GGdef}) for the determination of $\langle \frac{\alpha}{\pi}G\,G \rangle$.

Besides the determination of $\langle \frac{\alpha}{\pi}G\,G \rangle$ there is a general
interest in the behavior of perturbative series in QCD (for a recent investigation see~\cite{Meurice:2006cr}). 
Observable quantities can be written as series of the form
\begin{equation}
Q \sim\, \sum_{n} a_n \lambda^n\,,
\end{equation}
where $\lambda$ denotes some coupling. 
It is generally believed that these series are asymptotic, and
assumed that for large $n$ the leading growth of the coefficients $a_n$
can be parame\-trized as~\cite{ZinnJustin}
\begin{equation}
a_n \sim \,C_1\,(C_2)^n\,\Gamma (n+C_3)\,,
\label{coefffactorial}
\end{equation}
i.e., they show a factorial behavior. Using the technique of NSPT one reaches 
orders of the perturbative series where a possible set-in of this assumed
behavior can be tested. There is a recent paper
of Narison and Zakharov~\cite{Narison:2009ag} where the authors
discuss the difference between short and long perturbative series and
its impact on the determination of $\langle \frac{\alpha}{\pi}G\,G \rangle$.

In this paper we present perturbative calculations in NSPT up to order $n=20$ for
Wilson loops for lattice sizes $L^4$ with $L=4,\dots,12$. The
computation for $L=12$ were performed on a NEC SX-9 computer of RCNP at Osaka University,
all others on Linux/HP - clusters at  Leipzig University.
We calculate the Wilson loops in quenched QCD with plaquette gauge action.

\section{NSPT calculation up to $n=20$}

NSPT allows perturbative calculations on a lattice up to loop order $n$
which never will be reached by the standard diagrammatic approach.
The algorithm is introduced and discussed in detail in~\cite{Alfieri:2000ce, Di Renzo:2004ge} - 
we will not present it in this paper. We only want to point to some
essential topics:
\begin{itemize}
\item The computer implementation of NSPT requires the discretization of the so-called
(rescaled) Langevin time $\tau$
$$
\tau \rightarrow \tau_k= k \varepsilon/\beta\,, \quad k=0, 1, 2, \dots \,.
$$
($g^2=6/\beta$ is the bare lattice coupling). Practically,
this means that the corresponding quantities are measured for different small but finite
$\varepsilon$. 
The final result is obtained in the limit $\varepsilon \rightarrow 0$.
This must be done with great care in order to obtain reliable numeric results.
\item The connection to infinite volume is achieved by the limit $L \rightarrow \infty$
which requires an additional extrapolation of the corresponding finite $L$ results.
\end{itemize}

\begin{figure}[h!b!p!]
  \begin{center}
    \begin{tabular}{cc}
       \includegraphics[scale=0.63,clip=true]{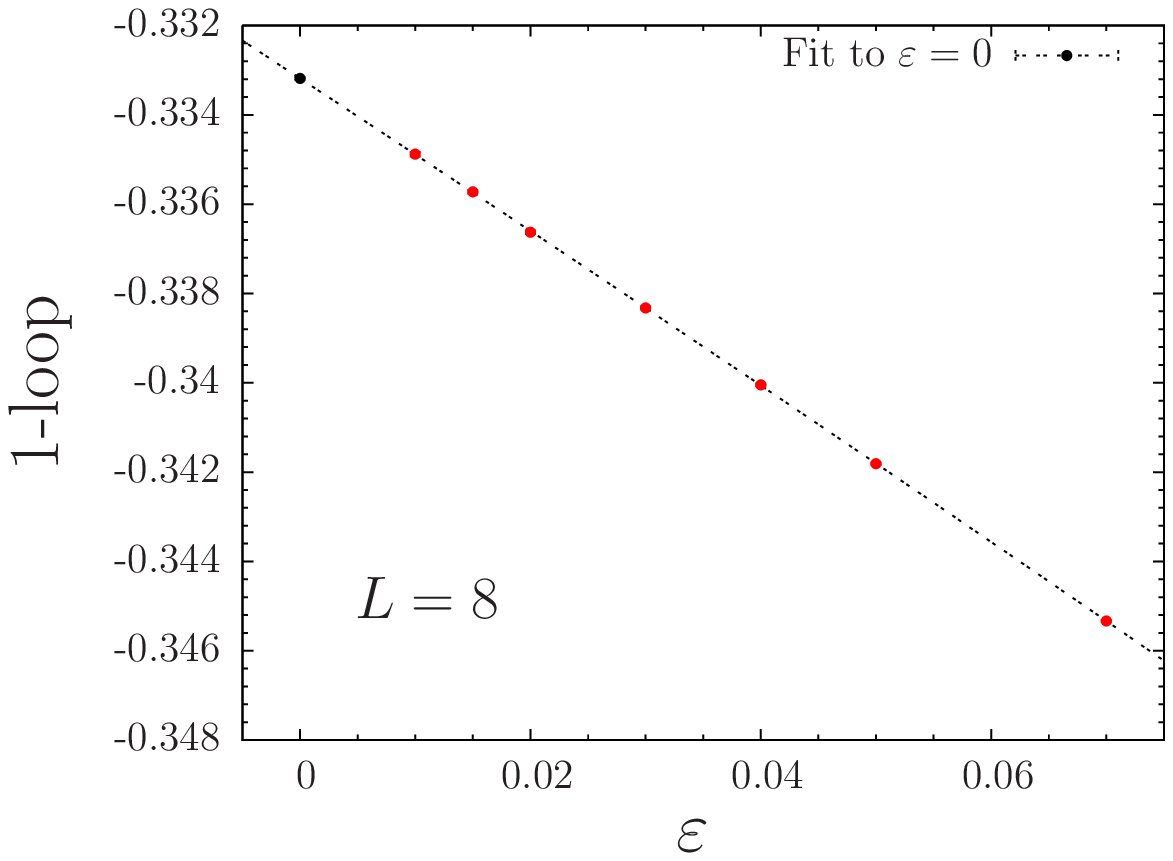}
&
       \includegraphics[scale=0.63,clip=true]{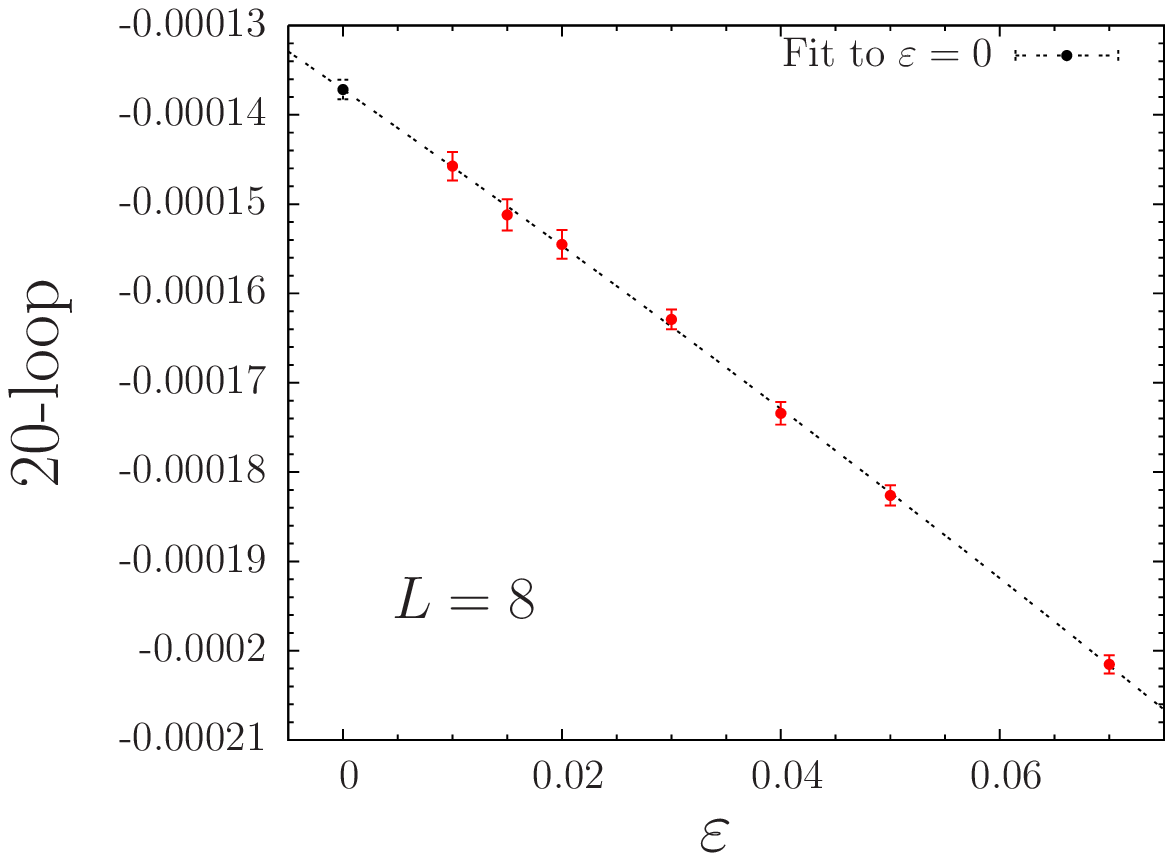}
    \end{tabular}
  \end{center}
\caption{Extrapolation $\varepsilon \rightarrow 0$  for $W_{11}$ for 1-loop (left) and 20-loop (right)
for $L=8$.}
\label{Figeps0}
\end{figure}

In Fig. \ref{Figeps0} we show the extrapolation $\varepsilon \rightarrow 0$ for 
lattice size $L=8$ for a plaquette, where we use a general quadratic ansatz in $\varepsilon$ for the fitting function.

\begin{figure}[htb!]
  \begin{center}
    \begin{tabular}{ll}
       \includegraphics[scale=0.6,clip=true]{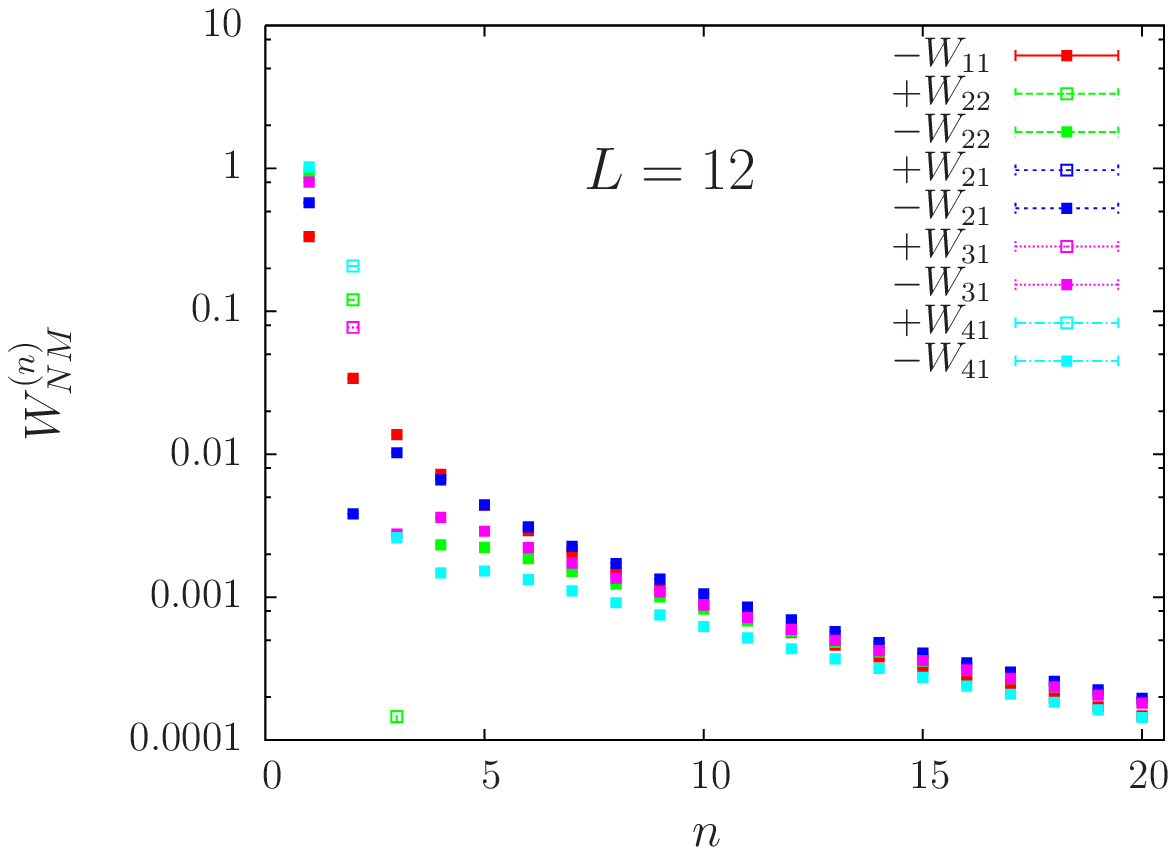} 
& 
       \includegraphics[scale=0.6,clip=true]{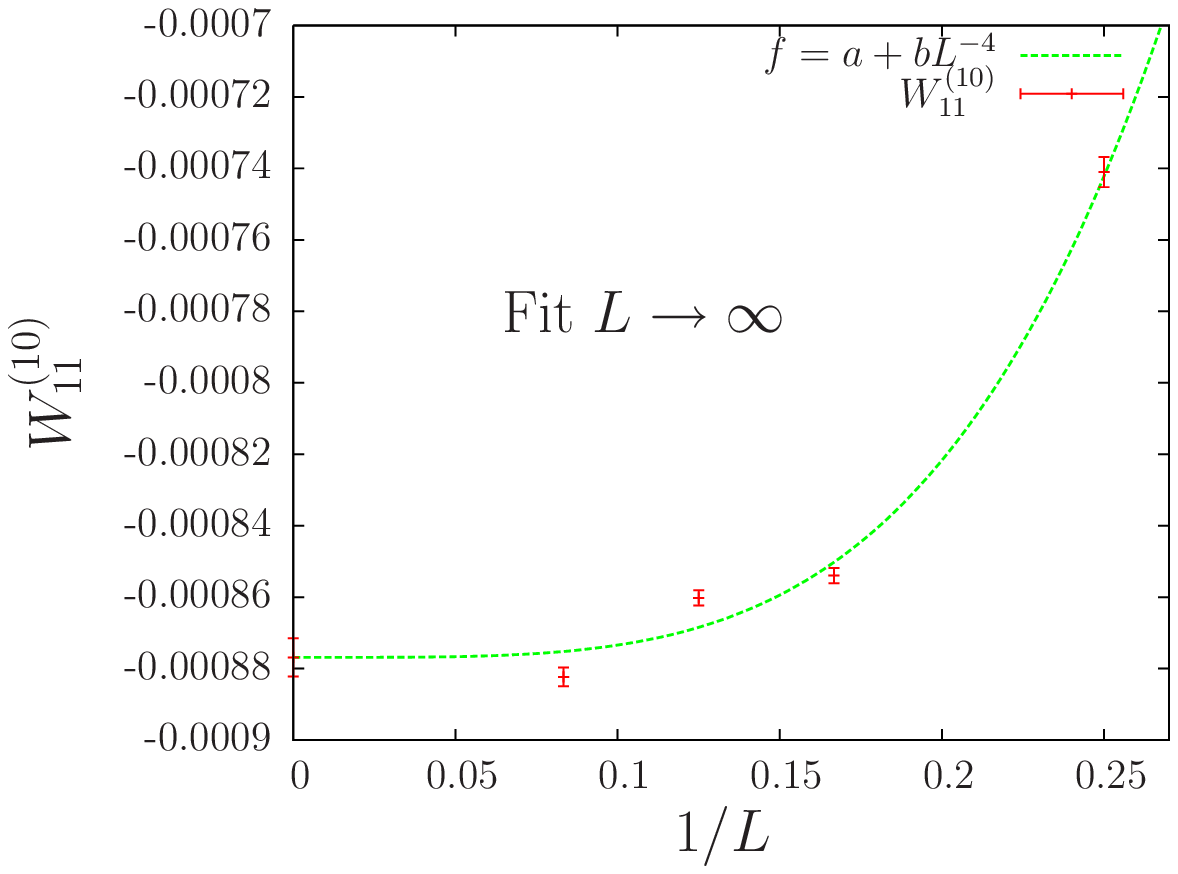} \\
    \end{tabular}
  \end{center}
\caption{Coefficients for various Wilson loops (left). Extrapolation $L \rightarrow \infty$ for $n=10$ (right).}
\label{Coffrat1}
\end{figure}

We write the general expansion of a Wilson loop of size $N\times M$ in terms of the
bare lattice coupling $g$ as
\begin{equation}
W_{NM} = \sum_{n=0}^{20}\,W_{NM}^{(n)}\, g^{2n}\,.
\end{equation}
Depending on the loop-size $(N,M)$  we found alternating signs for the 
perturbative coefficients
$W_{NM}^{(n)}$ for smaller $n$ whereas for larger $n$ they turn into a smooth asymptotic
behavior. An example is given in Fig. \ref{Coffrat1}  (left) for $L=12$.

A typical extrapolation to $L \rightarrow \infty$ for the plaquette is shown on the right side of Fig.
\ref{Coffrat1}.
Bali~\cite{Bali:2009} has computed one- and two-loop contributions to
Wilson loops of various sizes in the standard diagrammatic approach
for finite $L$. A comparison of our one- and two-loop NSPT results
with his results is given in Table \ref{BaliTab}. Based on the
results given by Bali we fixed the functional dependence of the
$L \rightarrow\infty$ extrapolation.
However, it should be empasized that this extrapolation becomes worse 
for larger loop sizes $(N,M)$.

\begin{table}[!htb]
\caption{Comparison of one- and two-loop results for NSPT and standard approach}
\vspace{0.5cm}
\centering
\begin{tabular} {c c r r r r}
\hline
$W_{NN}$ & $L$ & NSPT (1-loop) &  Bali (1-loop)  & NSPT (2-loop) &  Bali (2-loop) \\
 \hline
\\[1ex]
$W_{22}$ &  $4$   & $-0.87468(13)$   & $-0.87500$  & $0.10404(07)$    & $0.10406$\\[1.2ex]
         &  $6$   & $-0.90752(12)$   & $-0.90762$  & $0.11830(10)$    & $0.11837$\\[1.2ex]
         &  $8$   & $-0.91164(08)$   & $-0.91141$  & $0.12008(08)$    & $0.11993$\\[1.2ex]
         &  $12$  & $-0.91259(03)$   & $-0.91261$  & $0.12038(04)$    & $0.12038$\\[1.2ex]
$W_{33}$ &  $6$   & $-1.50088(30)$   & $-1.50093$  & $0.60906(34)$    & $0.60866$\\[1.2ex]
         &  $8$   & $-1.52873(23)$   & $-1.52803$  & $0.63693(23)$    & $0.63632$\\[1.2ex]
         &  $12$  & $-1.53526(47)$   & $-1.53533$  & $0.64370(13)$    & $0.64360$\\[1.2ex]
$W_{44}$ &  $8$   & $-2.14128(44)$   & $-2.14016$  & $1.52351(70)$    & $1.52331$\\[1.2ex]
         &  $12$  & $-2.16950(24)$   & $-2.16922$  & $1.57178(60)$    & $1.57006$\\[0.8ex]
 \\[0.5ex]
\hline
\label{BaliTab}
\end{tabular}
\end{table}

\section{Perturbative series at large order}

The order of perturbation theory we have reached in our calculations allows to study 
the large order behavior  and to test some models
concerning the $n-$dependence of the coefficients. 
This is essential in order to compute the perturbative part of the Wilson loops
as precise as possible.
In order not to interfere
with possible extrapolation ($L \rightarrow \infty$) effects we investigate
this for finite $L$.

\subsection{Heuristic model}

\begin{figure}[!htb]
  \begin{center}
   \includegraphics[scale=0.9,clip=true]{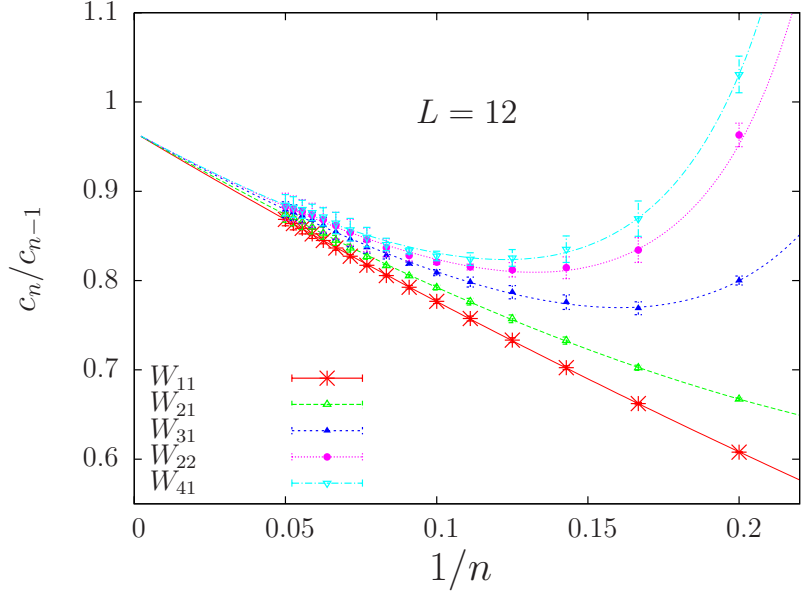}
  \end{center}
\caption{Domb-Sykes plot for various $W_{NM}$ together with their fits (3.3).}
\label{Coffrat}
\end{figure}

In~\cite{Horsley:2001uy} the authors propose to use a series expansion
for a quantity which shows a power-like singularity
\begin{equation}
W_{11} \sim (1-u \,g^2)^\gamma= \sum_n \, \frac{\Gamma(n-\gamma)}{\Gamma(n+1)\Gamma(-\gamma)}
\,(u\, g^2)^n = \sum_n \, c_n\,g^{2n}  \,.
\label{HRS}
\end{equation}
From (\ref{HRS}) one derives the ratio of
successive coefficients $c_n$ as (slightly modified by a parameter $s$ to account for a small curvature)
\begin{equation}
r_n=c_n/c_{n-1} = u \left(1-\frac{1+\gamma}{n+s}  \right)\,.
\label{HRSratio}
\end{equation}
In a Domb-Sykes plot - $r_n$ plotted against $1/n$ - this is almost a straight line.
In Fig. \ref{Coffrat} one observes that $r_n$ for $W_{11}$ follows  this simple functional form almost
ideally. 

However, the corresponding curves for larger Wilson loops of moderate size have a more pronounced
non-linear dependence on $1/n$ as can be seen in Fig. \ref{Coffrat}. 
This suggests to generalize ansatz (\ref{HRSratio}) by adding an extra power in $n$
(for a detailed discussion see~\cite{QCDSF:2009})
\begin{equation}
r_n=c_n/c_{n-1} = u\, \frac{n^2+(s-q-1)n+t}{n(n+s)}\,.
\label{HRSratio1}
\end{equation}
For $t=0$ relation (\ref{HRSratio1}) is identical to (\ref{HRSratio}). It gives a
hyperbola in a Domb-Sykes plot. In this paper we assume that the intercept $u$
has a universal value for all loop sizes $(N,M)$. It is determined
from $W_{11}$ which has been computed most precisely.
The other parameters $(q,s,t)$ depend on $(N,M)$. 
The corresponding curves are shown in Fig. \ref{Coffrat}. 
They are obtained from the fit ansatz (\ref{HRSratio1}) where the
parameters are determined in the interval $5 \le n \le 20$.
In this region the perturbative coefficients of the
considered Wilson loops show a common
asymptotic behaviour as can be seen in Figure \ref{Coffrat1}~(left).

\begin{figure}[!htb]
  \begin{center}
   \includegraphics[scale=0.8,clip=true]{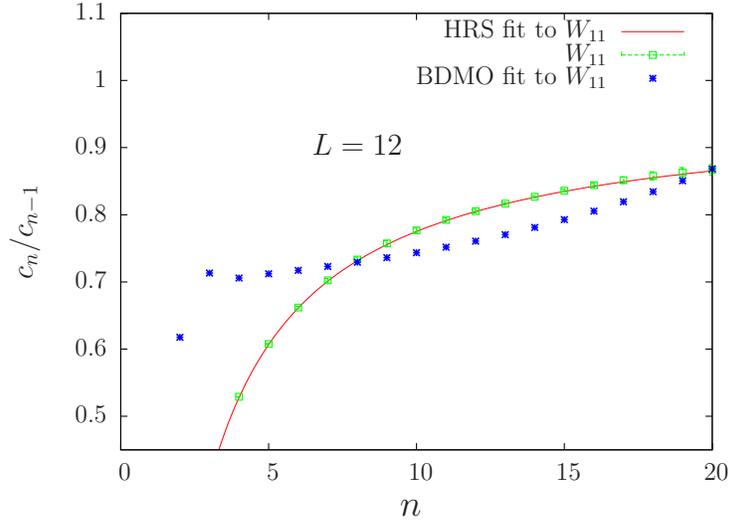}
  \end{center}
\caption{Comparison of $r_n$ of the plaquette $W_{11}$ for HRS and BDMO models}
\label{CoffratMod}
\end{figure}

There were speculations that already at order $n=10$ 
the perturbative coefficients show
a factorial growth due to renormalon contributions~\cite{Burgio:1997hc,DiRenzo:2000ua} (for a detailed
investigation of this point see also~\cite{Meurice:2006cr}).
For the plaquette  we plot in Fig. \ref{CoffratMod}
the ratio $r_n$ over $n$ for the ansatz (\ref{HRSratio}) (HRS) and the renormalon inspired model
as given in~\cite{Burgio:1997hc,DiRenzo:2000ua} (BDMO).
We do not observe a factorial growth, at least in the region
$n \le 20$ and for our lattice sizes.

\subsection{Boosted perturbation theory}

It is well-known that the bare lattice coupling $g$ is a bad expansion
parameter due to lattice artefacts like tadpoles.
There is a hope that by redefining the coupling $g$ into a
boosted coupling $g_b$ and the corresponding rearrangement of
the series a better convergence behaviour can be achieved.
For the plaquette $P=W_{11}$ we use the replacements
\begin{equation}
g^2 \rightarrow g^2_{b}=\frac{g^2}{P_{pert,b}}:\quad
P_{pert}(g,n^\star)=1+\sum_{n=1}^{n^\star}\,W_{11}^{(n)}\,g^{2n} 
\rightarrow P_{pert,b}(g_{b},n^\star)=1+\sum_{n=1}^{n^\star}\,W_{b,11}^{(n)}\,g^{2n}_{b}\,,
\label{BoostP}
\end{equation}
where $n^\star$ is the maximal loop order.

Boosted perturbation theory has been applied to improve the perturbative
series for the plaquette for the first time by Rakow~\cite{Rakow:2005yn}.
He showed that $P_{pert,b}(g_{b},n^\star)$ reaches a stable plateau
much earlier than $P_{pert}(g,n^\star)$ as a function of $n^\star$.
\begin{figure}[h!b!p!]
  \begin{center}
    \begin{tabular}{cc}
       \includegraphics[scale=0.6,clip=true]{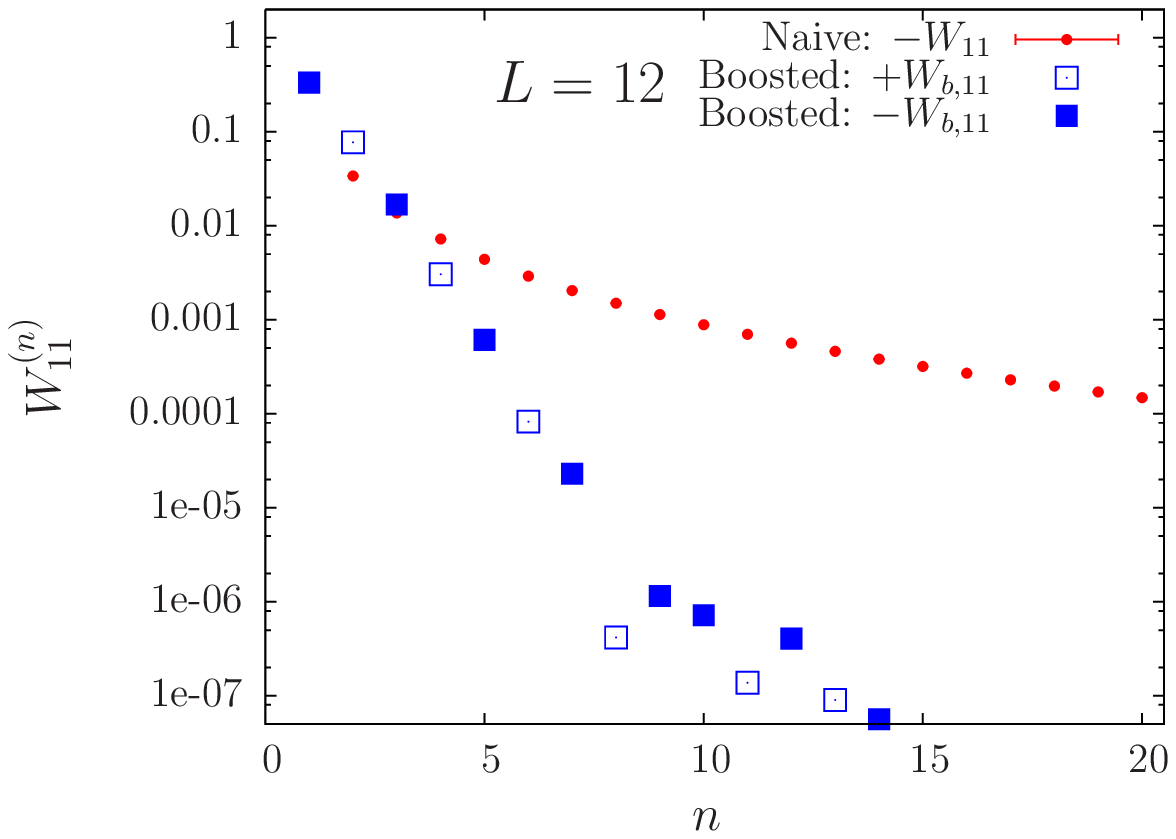} 
& 
       \includegraphics[scale=0.6,clip=true]{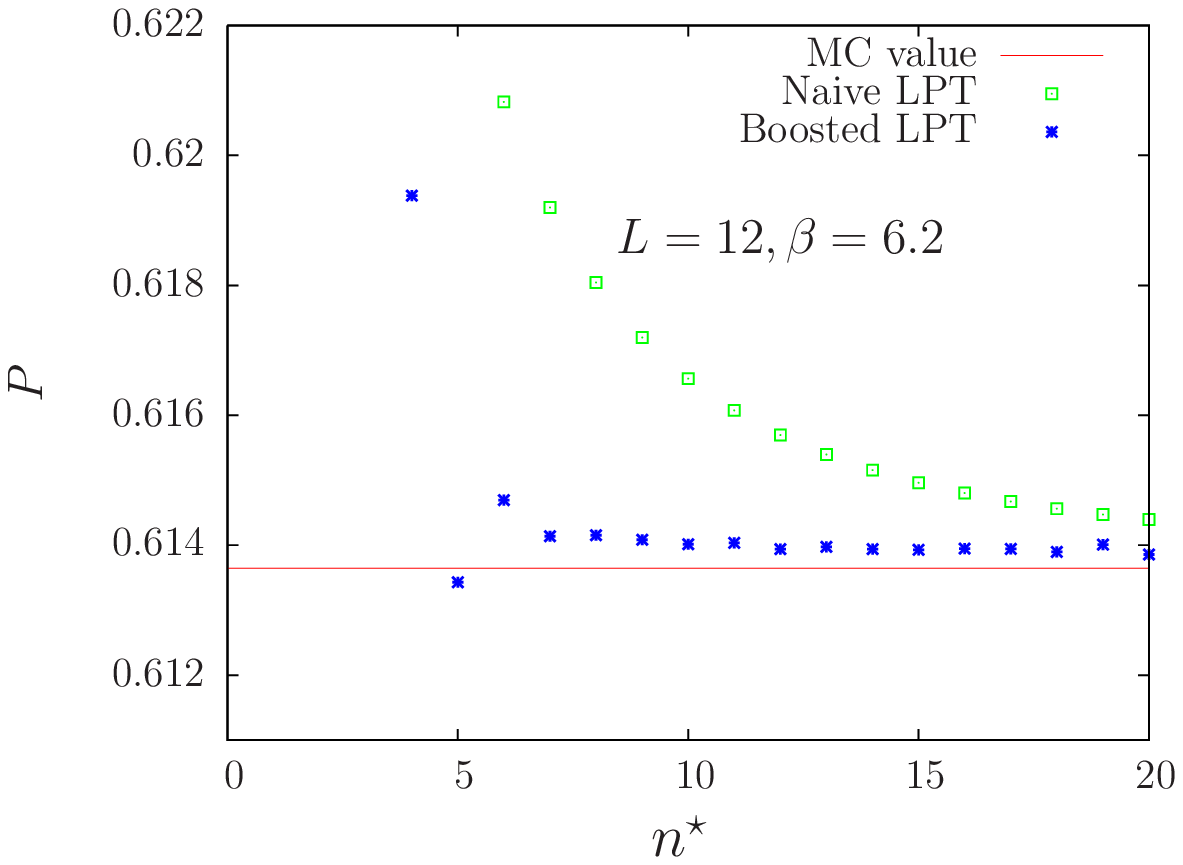} 
    \end{tabular}
  \end{center}
\caption{Coefficients for naive and boosted LPT (left). $P$ at $\beta=6.2$ as function of $n^\star$~(right).}
\label{Boost}
\end{figure}
Fig. \ref{Boost}~(left) shows that the boosted coefficients $W_{b,11}^{(n)}$ oscillate
but rapidly become very small. Of course, one should act with caution in the
region of $n$ where $|W_{b,11}^{(n)}| \sim 10^{-7}$. The superior convergence
behavior for the plaquette is demonstrated in Fig. \ref{Boost}~(right) confirming
the result in~~\cite{Rakow:2005yn}. The Monte Carlo result is taken from~\cite{Boyd:1996bx,Gockeler:2005rv}.

\section{Non-perturbative gluon condensate}

\begin{figure}[h!b!p!]
  \begin{center}
   \includegraphics[scale=0.7,clip=true]{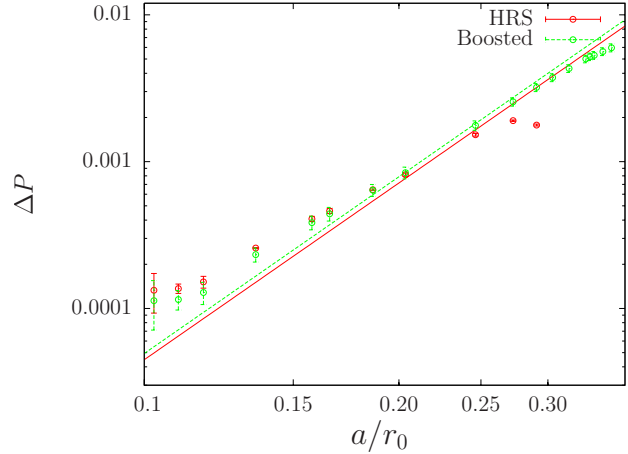}
  \end{center}
\caption{$\Delta P(a/r_0)$  with fit curves $\sim (a/r_0)^4$.}
\label{dPoverar0}
\end{figure}

As discussed in the introduction there are speculations whether the difference 
$\Delta P = P_{pert}-P_{MC}$ behaves as $\sim a^2$ or $\sim a^4$.
We can check this by plotting $\Delta P$ versus $a/r_0$ where $r_0$
denotes the Sommer scale.
The functional relation between $\beta$ and $r_0/a$ has been
taken from~\cite{Necco:2001xg}.
In Fig. \ref{dPoverar0} $\Delta P(a/r_0)$ is plotted in the infinite volume limit
($L \rightarrow \infty$) for
both models discussed in the previous sections. 
The MC data points have been taken from~\cite{Boyd:1996bx,Gockeler:2005rv}. 
(The cut-off in the
HRS-model data for larger $a$ is due to the convergence radius for the coupling determined
by the parameter $u$ in (\ref{HRS}).)
We make the ansatz $\Delta P(a/r_0)= C \, (a/r_0)^4$
and approximate $\left(\frac{-b_0\,g^2}{\beta(g)}\right) \sim 1 $.
This gives for the range $0.1 \le a/r_0 \le 0.25$
\begin{equation}
r_0^4\,\langle \frac{\alpha}{\pi}G\,G \rangle_{HRS} = 1.63(9), \quad
r_0^4\,\langle \frac{\alpha}{\pi}G\,G \rangle_{boosted} = 1.80(5).
\label{GGresr0}
\end{equation}
Fig. \ref{dPoverar0} shows that the data are well described by the
ansatz $\sim (a/r_0)^4$ over a large range of $a$.
Inserting , e.g. $r_0=0.5$ fm we obtain
\begin{equation}
\langle \frac{\alpha}{\pi}G\,G \rangle_{HRS} = 0.039(2)\, GeV^4, \quad
\langle \frac{\alpha}{\pi}G\,G \rangle_{boosted} = 0.043(2)\, GeV^4\, .
\label{GGres}
\end{equation}
One can try to fit the more general ansatz $\Delta P = C \, (a/r_0)^\delta$ to the
data. For the boosted model and $0.1 \le a/r_0 \le 0.25$ we get $\delta=3.5 \pm 0.1$
which is not too far from $\delta=4$.

All given errors are purely statistical, some of the systematic
 uncertainties are at least as large, and we are planning
 a more careful error analysis in the full paper~\cite{QCDSF:2009}.
It should be emphasized that the determination of the gluon condensate
depends less on the assumption of large loop order behavior than in
earlier investigations where all contributions beyond $n=10$ were
obtained by extrapolation.

\section{Summary}

In this paper we presented the perturbative calculation of Wilson loops of different
sizes up to loop order $n=20$ using NSPT. We compared three models to
describe the data: a renormalon inspired model (BDMO), a heuristic fit (HRS)
and boosted perturbation theory. We found that up to order $n=20$ the resulting curves
show a $\sim a^4$ behaviour. This supports the claim of Narison and Zakharov~\cite{Narison:2009ag}
that a behaviour $\sim a^2$ is due to perturbative series cut a lower order.
The values (\ref{GGres}) for $\langle \frac{\alpha}{\pi}G\,G \rangle$ found for HRS and boosted PT
are larger than obtained in other computations~\cite{Shifman:1978bx,Rakow:2005yn,Horsley:2001uy}.
The gluon condensate can also be obtained from larger and/or asymmetric Wilson loops
serving as an additional check. We hope to come back to this problem in~\cite{QCDSF:2009}.
\vskip 1cm
\leftline{\bf Acknowledgements}{\it This investigation has been supported partly by DFG. 
H.~P. acknowledges useful comments from Y. Meurice. P.~R. thanks S. Narison for 
helpful discussions.
We also thank the RCNP at Osaka university for using its NEC SX-9 computer.}

\end{document}